\documentclass[universe,article,accept,moreauthors]{Definitions/mdpi} 
\firstpage{1} 
\pubvolume{1}
\issuenum{1}
\articlenumber{0}
\pubyear{2026}
\copyrightyear{2026}
\externaleditor{Firstname Lastname} 
\datereceived{10 August 2026} 
\daterevised{9 October 2026} 
\dateaccepted{9 October 2026} 
\datepublished{ } 

\usepackage{graphicx,amsmath,booktabs,array,multirow,url}
\usepackage{balance}
\usepackage{bm}
\usepackage{xcolor}
\usepackage{newunicodechar}
\newunicodechar{″}{\arcsec}
\newunicodechar{′}{\prime}
\newunicodechar{α}{$\alpha$}
\newunicodechar{Λ}{$\Lambda$}
\newunicodechar{θ}{$\theta$}
\newunicodechar{κ}{$\kappa$}
\newunicodechar{β}{$\beta$}
\newunicodechar{Σ}{$\Sigma$}
\newunicodechar{✓}{\checkmark}
\newunicodechar{χ}{$\chi$}
\newcommand{\arcsec}{\ifmmode^{\prime\prime}\else''\fi}
\newcommand{\arcmin}{\ifmmode^{\prime}\else$'$\fi}

\usepackage[normalem]{ulem}
\newcommand{\rev}[1]{#1}                              
\newcommand{\del}[1]{}                                

\Title{{Gravitational} 
Lensing of the Bullet Cluster in the CCC+TL 
Cosmology\rev{: A Controlled Proof-of-Concept Study}}

\Author{{Rajendra P. Gupta} 
$^{1}$\orcidA{} and Nikolaos Samaras $^{2}$\orcidB{}}

\AuthorNames{Rajendra P. Gupta, and Nikolaos Samaras}

\address[1]{Department of Physics, University of Ottawa
, Ottawa, ON 
 K1N 6N5, Canada}

\corres{\hangafter=1 \hangindent=1.05em \hspace{-0.82em}Correspondence: rgupta4@uottawa.ca}

\abstract{\rev{We present a controlled proof-of-concept study of gravitational lensing in the Bullet Cluster (1E 0657-
56, $z_{d}=0.296$) under the Covarying Coupling Constants plus Tired Light (CCC+TL) cosmology.}~Using a JWST spectroscopic catalogue of 219 confirmed cluster members \rev{and the 135 gold multiple images in 49 systems from \citep{Rihtarsic2026} 
}, we construct independent convergence maps for CCC+TL and a simplified $\texorpdfstring{\Lambda}{Lambda}$CDM reference model using their respective angular-diameter distances and critical surface densities. The CCC+TL model adopts a two-parameter effective-density prescription \rev{anchored to the same virial-scale mass budget as the $\Lambda$CDM reference by construction, rather than fitted to} the multiple-image positions themselves. Three comparisons are presented: First, the CCC+TL and $\texorpdfstring{\Lambda}{Lambda}$CDM mean-convergence bands in radial rings overlap within the propagated \texorpdfstring{$\pm 30 \%$}{±30\%} virial-mass uncertainty across all rings containing multiple images (\texorpdfstring{$\theta\leq80\arcsec$}{theta<=80 arcsec}). {Secondly, on the source-plane back-projection diagnostic (convention $\alpha_{\mathrm{phys}}(z_{s})=\alpha_{\mathrm{dft}}\times\beta(z_{s})/\beta(z_{s,\mathrm{ref}})$ with $z_{s,\mathrm{ref}}=2$), the unoptimised two-parameter CCC+TL pipeline gives $\mathrm{RMS}=5.22\arcsec$ ($4.63\arcsec$--$5.91\arcsec$ across the mass band), compared with $6.82\arcsec$ ($5.75\arcsec$--$8.09\arcsec$) for the equivalent two-parameter $\Lambda$CDM reference. For context, the fully optimised 32-parameter Lenstool reconstruction of \citep{Rihtarsic2026} achieves an \emph{image-plane} RMS of $\sim\!0.41\arcsec$ on this gold catalogue; this is a different quantity than the source-plane values quoted above and is not directly comparable. Within the $\pm 30\%$ $M_{200}$ uncertainty band the two source-plane scatter ranges partially overlap, which is a statement about two model predictions, not a test of either model against the observed multiple-image positions.} \del{Secondly, on the source-plane scatter metric---the RMS spread of back-projected source positions within each multiple-image system---2-parameter CCC+TL yields $7.36\arcsec$, compared with $6.56\arcsec$ for the equivalent $\Lambda$CDM reference model (2-parameter unoptimised, rather than 32-parameter optimised), a difference covered by the same mass uncertainty.}~Thirdly, the CCC+TL effective extra-gravity distribution follows the galaxy distribution rather than the X-ray gas, reproducing the qualitative morphological feature that defines the Bullet Cluster\rev{; we stress that this agreement is partly structural, since the galaxy-scale and cluster-scale CCC profiles are centred on the member galaxies by construction}. For the same cluster mass, the Einstein-radius scaling in arcseconds differs by only 1.7\% between the two cosmologies. \del{These results show that, within the present controlled comparison and current mass uncertainties, the Bullet Cluster strong-lensing data do not exclude the CCC+TL framework.} \rev{These results are a necessary first step toward a full image-plane likelihood test of CCC+TL on the Bullet Cluster, but not a sufficient one; the full pipeline reproducing the numerical results of this paper is available at a dedicated website.}}


\begin{document}

\section{Introduction}\label{introduction}

Gravitational lensing provides one of the most direct probes of mass in
the Universe because it is sensitive to the total projected gravitating
matter along the line of sight, independent of its dynamical or
radiative state.~In the weak-lensing regime, coherent distortions of
background galaxy shapes trace the projected surface density of
foreground structures, while in the strong-lensing regime multiple
images, giant arcs and Einstein rings constrain the inner projected
potential with high precision. For galaxy clusters, the combination of
strong and weak lensing has therefore become one of the principal routes
to reconstructing two-dimensional mass distributions and testing the
relation between baryons and the dominant gravitating component
\citep{Narayan1996,Bartelmann2001,Umetsu2020,Cho2025}.

Cluster lensing has played a particularly important role in the modern
dark-matter paradigm.~The Bullet Cluster, 1E 0657-56, remains the
best-known example because its merger geometry produces a spatial
separation between the collisional X-ray-emitting gas and the
approximately collisionless galaxy component. Chandra observations
identified a bow shock with Mach number of order 2--3, implying a
recent, nearly plane-of-sky merger, while weak- and strong-lensing
reconstructions showed that the main lensing peaks are displaced from
the dominant baryonic gas component and lie much closer to the galaxy
concentrations. These results established the Bullet Cluster as a
stringent empirical test for any theory seeking to explain cluster
lensing without invoking conventional non-baryonic dark matter
\citep{Markevitch2002,Clowe2004,Markevitch2006,Clowe2006,Bradac2006}.

Recent JWST observations have sharpened this test considerably. Using
JWST/NIRCam imaging and NIRSpec spectroscopy, \citet{Rihtarsic2026}
constructed an updated Bullet Cluster lens model constrained by 135
secure multiple images from 27 background galaxies with spectroscopic
redshifts, and explicitly modelled the system with several large-scale
haloes, cluster members, intracluster gas and surrounding group-scale
haloes using Lenstool. These data now make it possible to revisit
cluster lensing with alternative mass prescriptions while holding the
observational constraints fixed \citep{Rihtarsic2026,Niemiec2020}.

The present work is motivated by the Covarying Coupling Constants (CCC)
framework and its extension with Tired Light (TL), CCC+TL, in which dark
matter and dark energy are interpreted not as independent substances but
as emergent effects of a modified cosmological framework. In earlier
work, this programme was developed at the background level and then
applied to galaxy rotation curves, cosmic expansion tests and
cluster-scale phenomenology, including the suggestion that the apparent
dark-matter component may arise from a density-dependent effective
gravitating response rather than from a separate particle species
\citep{Gupta2023,Gupta2024a,Gupta2024b,Gupta2025,Gupta2026}.

A particularly useful feature of the CCC approach is that it naturally
suggests a distinction between dense baryonic cores and lower-density
outer regions. In the application developed here, the baryonic
distribution of each cluster member galaxy is represented by a Hernquist
profile, which is well suited to spheroidal stellar systems and has the
practical advantage that both its three-dimensional density and
projected quantities are analytically tractable in the standard case
\citep{Hernquist1990}.~The CCC prescription is then introduced beyond a
characteristic turn-off radius, where the effective gravitating density
is taken to depart from the pure baryonic Hernquist form. This
construction leads to an enhanced projected surface density without
altering the basic lensing formalism itself.

The purpose of this paper is not to modify gravitational lensing theory
itself, but to test whether the observed Bullet Cluster strong-lensing
signal can be reproduced when the usual galaxy-scale halo prescription
is replaced by a CCC effective density profile tied to the observed
member-galaxy positions, fluxes and derived structural scales. This
makes the comparison especially transparent: the redshifts,
multiple-image constraints, gas component and lensing geometry remain
standard, while only the assumed galaxy-member mass prescription is
altered.

\textls[-5]{Accordingly, the present analysis should be viewed as a controlled
substitution test rather than as a full optimised lens-model
competition. We ask whether a CCC+TL effective-density prescription,
applied to the same member catalogue and observational inputs, can
generate convergence maps, enclosed projected masses and source-plane
scatter values consistent with those of a simplified $\Lambda$CDM reference
model within current mass uncertainties.}

We therefore determine whether the CCC+TL model, applied consistently to
the 219-member JWST galaxy catalogue of \citet{Rihtarsic2026} and
without fitting to the multiple-image positions, produces a mass
distribution compatible with the observed strong-lensing signal. Section \ref{the-ccctl-cosmological-model} presents the theoretical framework, Section \ref{data-and-pipeline} describes the data and
pipeline, Section \ref{results} presents the results, Section \ref{discussion} discusses their
implications and limitations, and Section  \ref{conclusions} summarises the conclusions.


\subsection{The CCC+TL Cosmological Model}\label{the-ccctl-cosmological-model}

Fundamentals of the CCC and CCC+TL model generating extra gravity,
resulting in illusionary \(\alpha\)-matter, have been presented,
discussed, and applied to galaxy rotation curves, cluster observations,
and gravitational lensing in earlier papers \citep{Gupta2025,Gupta2026}.
Therefore, we include here only what has direct relevance to this work.

A key structural ingredient of CCC is the invariant ratio
\(G/c^{3}\  = \gamma\), motivated by the form of the Einstein--Hilbert
action, which requires the prefactor \(c^{3}/G\) to be time-independent
if the gravitational field equations are to retain their standard form.~This motivates a scaling ansatz in which only quantities with non-zero
length dimensionality vary: if a quantity \(X\) has length dimension
\(L^{n}\), then \(X(t) = X(0)f^{n}(t)\) for some dimensionless
microstructural scaling function \(f(t)\).~A simple realisation is
\(c(t) = c(0)f(t)\), \(G(t) = G(0)f^{3}(t)\),
\(\hslash(t) = \hslash(0)f^{2}(t)\), and \(k_{B} = k_{B}(0)f^{2}(t)\),
so that \(G/c^{3}\) is constant, and Planck units acquire a natural
microscopic interpretation. The length dimensionality rule is modified
slightly for the expressions involving time differentiation or
integration, such as for the Hubble rate with no length dimension
scaling as \({f(t)}^{- 1}\).

An easily workable function is
\(f(t) = \exp\left( \alpha\left( t - t_{0} \right) \right)\) with
\(\alpha\) a free parameter.~The Friedmann equations then acquire
additional terms that behave effectively as virtual dark energy
(\(\alpha\)-energy) and virtual dark matter (\(\alpha\)-matter) without
invoking new particles \citep{Gupta2023,Gupta2025}.~The parameter
\(\alpha\) may be determined by fitting observational data, such as
supernovae type 1A (Pantheon+) or CMB (Planck) data.~The angular
diameter distances \(D_{A}(z_{1},\ z_{2})\) in CCC+TL substantially
exceed their $\Lambda$CDM counterparts, particularly at high \mbox{source~redshift}.

In the present work, the CCC+TL framework enters only through (i) the
effective density prescription used for the galaxy and cluster
components, and (ii) the cosmology-dependent angular-diameter distances
entering \(\Sigma_{crit}\).

\subsection{Gravitational Lensing}\label{gravitational-lensing}

In the thin-lens approximation, the observable quantity relevant for
lensing is the projected surface mass density at angular vector
coordinates \(\theta\) and radial coordinate 
\(z\),
\begin{equation}
\Sigma\left( \theta \right) = \int \rho(D_{d}\theta, z)\,dz
\end{equation}
where \(\rho\) is the density and \(D_{d}\) is the
angular-diameter distance to the deflector (lens). The dimensionless
convergence is then
\begin{equation}
\kappa(\theta) = \frac{\Sigma(\theta)}{\Sigma_{crit}}
\end{equation}
with critical surface density \vspace{+6pt}
\begin{equation}
\Sigma_{crit} = \frac{c^{2}}{4\pi G}\frac{D_{s}}{D_{d}D_{ds}}
\end{equation}

Here, \(D_{d} = D_{A}\left( 0,\ z_{d} \right)\),
\(D_{s} = D_{A}(0,\ z_{s})\), and \(D_{ds} = D(z_{d},\ z_{s})\) are the
angular-diameter distances to the deflector at redshift \(z_{d}\), to
the source at redshift \(z_{s}\), and from deflector to source,
respectively, with the observer at redshift 0. These are geometric
lensing factors rather than Euclidean separations, and
\(D_{ds}\) must be computed independently from the
cosmological distance relation rather than approximated by
\(D_{s} - D_{d}\) \citep{Narayan1996,Bartelmann2001,Umetsu2020}. For a
flat background, one may write
\(D_{ds} = \left\lbrack \left( 1 + z_{s} \right)D_{s} - \left( 1 + z_{d} \right)D_{d} \right\rbrack/(1 + z_{s})\).

For an axisymmetric lens, the projected mass interior to projected
radius \(R\) is
\begin{equation}
M_{2D}(<R) = 2\pi\int_{0}^{R} \Sigma\left( R' \right) R'\,dR'
\end{equation}
and the corresponding deflection angle is proportional to
\(M_{2D}( < R)/R\).~In practice, standard cluster lens models represent
the mass distribution as a superposition of large-scale smooth haloes,
galaxy-member sub haloes and, when data permit, an explicit gas
component.~Modern implementations combine strong-lensing constraints in
the core with weak-lensing information at larger radii, and public
software such as Lenstool now supports such joint modelling in a
self-consistent framework \citep{Bradac2006,Niemiec2020,Rihtarsic2026}.

In the present CCC implementation, we retain this standard lensing
framework and alter only the mass prescription assigned to the
cluster-member galaxies. The baryonic density of each member galaxy is
taken to follow a Hernquist profile,
\begin{equation}
\rho_{b,i}(r) = \frac{M_{b,i}}{2\pi}\frac{r_{a,i}}{r\left( r + r_{a,i} \right)^{3}}
\end{equation}
where \(M_{b,i}\) is the baryonic, or in practice stellar,
mass assigned to galaxy \(i\), and \(r_{a,i}\) is its characteristic
radius \citep{Hernquist1990}.

Following the CCC prescription used in our earlier work
\citep{Gupta2025,Gupta2026}, we define a turn-off radius
\(R_{t,i} = r_{a,i}\), and hence a turn-off density
\begin{equation}
\rho_{t,i} = \rho_{b,i}\left( r_{a,i} \right) = \frac{M_{b,i}}{16\pi r_{a,i}^{3}}
\end{equation}

For \(r \leq r_{a,i}\), the effective gravitating density is
taken to be purely baryonic,
\begin{equation}
\rho_{eff,i}(r) = \rho_{b,i}(r)
\end{equation}

For \(r > r_{a,i}\), the effective gravitating density is
replaced by the CCC form
\begin{equation}
\rho_{eff,i}(r) \equiv \rho_{CCC,i}(r) = 2\sqrt{2}\rho_{t,i}
\left[ \frac{r}{r_{a,i}} \left( 1 + \frac{r}{r_{a,i}} \right)^{3} \right]^{-1/2}
\end{equation}

While \(\rho_{t,i}\) and \(r_{a,i}\) are expected to be galaxy specific
due primarily to their varying morphologies, in practice, we define a
single turnoff density \(\rho_{t}\) for all galaxies and determine
\(r_{a,i}\) for each galaxy of stellar mass \(M_{b,i}\) using Equation (6).
This computation simplification eliminates the need to know the
characteristic radius of each galaxy at the cost of losing details of
the lensing profile. This construction preserves the baryonic inner
profile while generating a more slowly declining effective outer density
derived from CCC, thereby enhancing the projected lensing signal without
introducing an explicit particle-dark-matter halo. The physical
motivation and the associated density-threshold picture were developed
in \citep{Gupta2025,Gupta2026}.

For the cluster-scale component, a pure Hernquist form is too centrally
concentrated for the present application. We therefore adopt a
phenomenological modified-Hernquist profile with an
\(r^{- 1}\left( r + r_{a} \right)^{- 2}\) dependence, chosen as a simple
spherical ansatz that is shallower in the outskirts and qualitatively
closer to NFW-like behaviour. We emphasise that this is an empirical
cluster-scale prescription, universality of which is yet to be verified:
\begin{equation}
\rho_{b,clust}(r) = \frac{M_{b,clust}}{2\pi}\frac{1}{r\left( r + r_{a,clust} \right)^{2}}
\end{equation}

Using the prescription similar to galaxies above, we define for clusters
turn-off radius \(R_{t,clust} = r_{a,clust}\) and turn-off density
\(\rho_{t,clust} = \rho_{b,clust}\left( r_{a,clust} \right) = \frac{M_{b,clust}}{(8\pi r_{a,clust}^{3})}\).

As is the case of galaxies, for \(r \leq r_{a,clust}\), the
effective gravitating density for clusters is taken to be purely
baryonic, \vspace{+6pt}
\begin{equation}
\rho_{eff,clust}(r) = \rho_{b,clust}(r)
\end{equation}

For \(r > r_{a,clust}\), the effective gravitating density is
replaced by the CCC form
\begin{equation}
\begin{split}
\rho_{eff,clust}(r) &\equiv \rho_{CCC,clust}(r) \\
&= 2\rho_{t,clust}\left[\frac{r}{r_{a,clust}}\left(1 + \frac{r}{r_{a,clust}}\right)^{2}\right]^{-1/2}
\end{split}
\end{equation}

As we will see below, the turn-off density at the galaxy scale is
substantially higher than at the cluster scale.

Now we can consider the projection of the density profiles to lensing
quantities. For a spherically symmetric galaxy profile, the projected
surface density of each galaxy is obtained via the Abel integral,
\begin{equation}
\Sigma_{i}(R) = 2\int_{R}^{\infty} \frac{\rho_{eff,i}(r)\,r\,dr}{\sqrt{r^{2} - R^{2}}}
\end{equation}

Here, for a galaxy centred at \((x_{i},\ y_{i})\),
\begin{equation}
R_{i}(x,y) = \sqrt{\left( x - x_{i} \right)^{2} + \left( y - y_{i} \right)^{2}}
\end{equation}

Then the total galaxy contribution on the cluster map is
\begin{equation}
\Sigma_{gal}(x,y) = \sum_{i} \Sigma_{i}\left( R_{i}(x,y) \right)
\end{equation}

When the gas map is available, it is added as a separate layer:
\begin{equation}
\Sigma_{tot}(x,y) = \Sigma_{gal}(x,y) + \Sigma_{gas}(x,y)
\end{equation}

The convergence is defined as
\begin{equation}
\kappa(x,y) = \frac{\Sigma_{tot}(x,y)}{\Sigma_{crit}}
\end{equation}

Because the lensing formalism remains unchanged, the test of the CCC
prescription is reduced to a comparison of the resulting convergence
field and image-plane predictions against those obtained from the
standard parametric halo model.

The convergence test used in this work is, therefore, straightforward.
First, a baseline lens model is constructed using the standard
flux-scaled galaxy prescription employed in the observational study.
Secondly, the galaxy-member contribution is replaced by the CCC
effective profile while the observational constraints, redshifts, gas
component, and lensing geometry are kept fixed. The resulting
\(\kappa(x,y)\), aperture masses and, where relevant, image residuals
can then be compared directly. If the CCC prescription is viable, it
should reproduce the principal convergence peaks and lensing strengths
without requiring the same galaxy-associated dark-halo component as in
the standard model.~This type of controlled substitution is especially
natural for the recent Bullet Cluster JWST analysis, where the member
catalogue and F277W-based scaling information are already available from
the modelling framework of \citet{Rihtarsic2026}.

At the Bullet Cluster geometry (\(z_{d} = \ 0.296\), representative
source redshift \(z_{s}\  = \ 2.0\)), the key quantities are compared in
Table~\ref{tab:distances}.

\begin{table}[H]
\centering
\caption{Cosmological distances and critical surface densities at the Bullet Cluster geometry (\mbox{$z_d=0.296$}, $z_s=2.0$). The CCC+TL and $\Lambda$CDM distances are determined following \citet{Gupta2023}.}
\label{tab:distances}
\begin{tabularx}{\textwidth}{lCCc}
\toprule
\textbf{Quantity} & \textbf{\boldmath{$\Lambda$}CDM} & \textbf{CCC+TL} & \textbf{(CCC+TL/\boldmath{$\Lambda$}CDM)} \\
\midrule
$D_d$ (Mpc) & 910.3 & 1090.3 & 1.198 \\
$D_s$ (Mpc) & 1726.7 & 4408.4 & 2.553 \\
$D_{ls}$ (Mpc) & 1333.4 & 3937.4 & 2.953 \\
$D_{ls}/D_s$ & 0.7723 & 0.8932 & 1.156 \\
$\Sigma_{\rm crit}$ ($M_\odot\,\mathrm{kpc}^{-2}$) & $2.365\times10^{9}$ & $1.707\times10^{9}$ & 0.722 \\
kpc per arcsec & 4.4131 & 5.2859 & 1.198 \\
\bottomrule
\end{tabularx}
\end{table}

The CCC+TL (CTL) distances are larger
than \(\Lambda\)CDM (LC) by a factor of
\(F \equiv \frac{D_{A}^{CTL}}{D_{A}^{LC}}\), so all physical lengths in
the CCC+TL pipeline scale as \(F \times\) their \(\Lambda\)CDM values.~The critical surface density,
\(\Sigma_{crit} = (\frac{c^{2}}{4\pi G})(\frac{D_{s}}{D_{d}D_{ds}})\),
is 27.8\% lower in CCC+TL than in $\Lambda$CDM at \(z_{s} = 2\). These two
effects---lower \(\Sigma_{crit}\) and larger physical scales
(\(F^{2}\) suppression of projected surface density)---almost exactly
cancel, as shown in Section~\ref{einstein-radius-scaling-between-cosmologies}.

Hereafter, since the density prescription described above is an integral
component of the CCC+TL model, we refer to all derived quantities---effective density profiles, $\alpha$-matter distributions, and
convergence maps---simply as CCC+TL (abbreviated CTL in figures and
tables), without further reference to CCC alone unless necessary.

\subsection{Two-Parameter CCC+TL Solution}\label{two-parameter-ccctl-solution}

The CCC+TL implementation used here contains two calibration parameters.
These are not fitted to the multiple-image positions, but are fixed by
requiring consistency with the adopted virial-scale mass budget of the
cluster.

\begin{enumerate}[label=,leftmargin=0em]
\item \textbf{Galaxy-scale turn-off 
\boldmath{\((\eta_{\rho}^{gal})\)}:} The turn-off radius of
each galaxy follows the density prescription \citep{Gupta2026} \vspace{+6pt}
\begin{equation}
r_{a} = \left( \frac{M_{*}}{16\pi\eta_{\rho}^{gal}\rho_{ref}} \right)^{1/3}
\end{equation}
where \(\rho_{ref} = 4.2 \times 10^{- 24}\ \text{g}\ \text{cm}^{- 3}\) is 
the reference density calibrated from galaxy rotation curves, an average of all seven
galaxies of different morphologies that \citet{Gupta2025} studied. The
galaxy-scale normalization parameter \(\eta_{\rho}^{gal}\) is calibrated
such that the total CCC+TL projected mass at the virial scale matches
the adopted reference-model mass budget, i.e., the CCC+TL and reference
$\Lambda$CDM pipelines are normalised to have the same total projected mass
within \(r_{200}\). Notably, the projected mass inside a finite aperture
is not the same as the total 3D virial mass within \(r_{200}\).
\end{enumerate}

The present work does not establish a universal value of
\(\eta_{\rho}^{gal}\); rather, it tests whether values consistent with
earlier galaxy-scale work can also reproduce the Bullet Cluster lensing
signal.

\begin{enumerate}[label=,leftmargin=0em]
\item \textbf{Cluster-scale turn-off \boldmath{\((\eta_{\rho}^{clus})\)}:} The
cluster-scale \(\alpha\)-matter is seeded by the total cluster stellar
mass, \(M_{*}^{clus} = f_{*}M_{200}\), where \(f_{*} = 0.015\) is the
stellar mass fraction \citep{Paraficz2016}. Specifically, \(M_{200}\) is
the total mass enclosed within a sphere of radius \(r_{200}\), within
which the mean mass density is exactly 200 times the critical density of
the universe at the cluster's redshift. Its turn-off parameter is set by
requiring the cluster-scale \(\alpha\)-matter mass to equal the
$\Lambda$CDM dark matter mass within \(r_{200}\):
\begin{equation}
\eta_{\rho}^{clus}\int_{0}^{r_{200}} 4\pi r^{2}\rho_{eff}^{clus}(r)\,dr = \left( 1 - f_{gas} - f_{*} \right) M_{200}
\end{equation}
where \(f_{gas}\  = \ 0.12\) is the gas fraction.~This gives
\(\eta_{\rho}^{clus}\  = \ 0.0134\), corresponding to a turn-off radius
\(r_{a}^{clus} = 102.6\) kpc for the main cluster
(\(M_{200} = 1.5\  \times \ 10^{15}M_{\odot}\)) calculated using
\(r_{a}^{3} = \frac{M_{*}^{clus}}{(8\pi\eta_{\rho}^{clus}\rho_{ref})}\).
\end{enumerate}

Both parameters are derived without reference to the multiple-image
positions, so the CCC+TL pipeline has \(k = 2\) effective degrees of
freedom.

We do not assume that the Bullet Cluster is dynamically relaxed or in
hydrostatic equilibrium. The use of virial-scale quantities such as
$M_{200}$ serves here only as a normalization convention for the
projected lensing comparison, while the actual observable test is
carried by the convergence field and source-plane scatter.

\subsection{Einstein Radius Scaling Between Cosmologies}\label{einstein-radius-scaling-between-cosmologies}

For an axisymmetric lens, the Einstein radius satisfies
\(\overline{\kappa}\left( < \theta_{E} \right) = 1,\ \) or equivalently
\(M_{2D}\left( < \theta_{E} \right) = \pi\left( D_{d}\theta_{E} \right)^{2}\Sigma_{crit}\).
For the same physical mass distribution expressed in angular units, the
change in physical scale per arcsecond between the two cosmologies
modifies the projected surface density by a factor \(F^{- 2}\), while
the lensing threshold changes through \(\Sigma_{crit}\).~Hence for the
same cluster mass \(M_{200}\) the Einstein radius in arcseconds rescales
approximately as (CTL is abbreviation for CCC+TL and LC is for
\(\Lambda\)CDM)
\begin{equation}
\theta_{E}^{CTL} = \theta_{E}^{LC} \times \sqrt{\frac{\Sigma_{crit}^{LC}}{F^{2}\Sigma_{crit}^{CTL}}}
\end{equation}

Substituting the values of Table~\ref{tab:distances}:
\begin{equation}
\frac{\theta_{E}^{CTL}}{\theta_{E}^{LC}} = \sqrt{\frac{2.365 \times 10^{9}}{{1.198}^{2} \times 1.707 \times 10^{9}}} = 0.983
\end{equation}

The lower \(\Sigma_{crit}\) in CCC+TL (which enlarges \(\theta_{E}\))
is almost exactly cancelled by the larger physical scales
(\(F^{2}  =  1.435\)), giving a net difference of only 1.7\%.~This near- perfect cancellation is not a coincidence: it reflects the
same compensation found for the convergence itself
(\(\kappa_{net}  \propto \lbrack\Sigma_{crit}^{LC}\ /\Sigma_{crit}^{CTL}\rbrack/F^{2}  =  0.966\)),
and means that \emph{for the same mass, CCC+TL and $\Lambda$CDM
predict essentially identical Einstein radii in arc-seconds.}

The $\pm$30\% uncertainty in \(M_{200}\) propagates as \(\theta_{E}\  \propto \sqrt{M}\), giving a \(\approx \pm 15\%\) spread in \(\theta_{E}\) that is common to both models.

\textls[-15]{Our framework follows the observational decomposition adopted by
\mbox{\citet{Rihtarsic2026}}, who model the Bullet Cluster with multiple
large-scale haloes, cluster members, intracluster gas, and surrounding
group-scale structure.~The present work does not attempt to reproduce
that full optimised model, but instead performs a controlled
substitution test in which the standard galaxy-halo prescription is
replaced by a CCC+TL effective-density prescription while retaining the
same observational inputs as far as possible. Group-scale substructure
and full dynamical modelling remain beyond the scope of the present
pilot calculation.}

\section{Data and Pipeline}\label{data-and-pipeline}

\subsection{Observational Data}\label{observational-data}

We use three observational inputs from \citet{Rihtarsic2026}. \enlargethispage{0.3cm}

\begin{figure}[H]
\includegraphics[width=0.86\columnwidth]{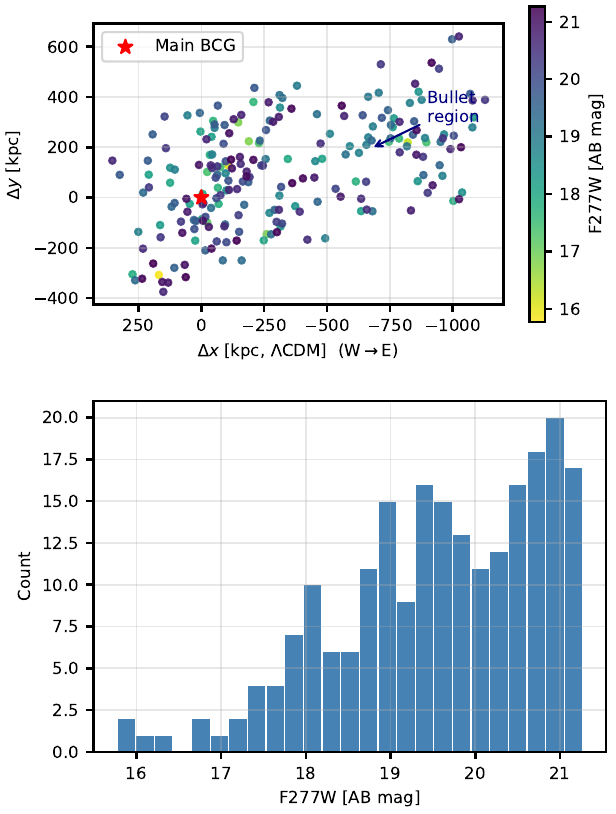}
\caption{JWST member-galaxy catalogue. (\textbf{Top}): projected sky positions of the 219 JWST spectroscopically confirmed cluster members relative to the main cluster BCG (red star), \mbox{colour-coded by F277W}}
\label{fig:catalogue}
\end{figure}

\begin{enumerate}[labelsep=5mm]
\item[1.] \textbf{Member-galaxy catalogue.} This comprises 219 spectroscopically
confirmed cluster members with ICRS (International Celestial Reference
System) coordinates, F277W AB magnitudes
(\(m_{277} = 15.77 - 21.27\) mag), and spectroscopic redshifts.
Stellar masses are derived from the calibration of Rihtar\v{s}i\v{c} et al.,
\begin{equation}
\log_{10}{{(M}_{*}/M_{\odot})} = - 0.406\ (m_{277} - 20) + 9.954,
\end{equation}
giving
\(M_{*} = 2.7\  \times \ 10^{9} - 4.7\  \times \ 10^{11}\ M_{\odot}\).
A physical scale of \(1\arcsec = 4.413\) kpc is used for \(\Lambda\)CDM
\citep{Rihtarsic2026}; the CCC+TL pipeline uses $1\arcsec = 5.286$ kpc.

\item[2.] \textbf{Gold multiple-image catalogue.} This comprises 135 secure multiple images
from 49 lensed systems (27 distinct galaxies), all with spectroscopic
redshifts spanning \(z_{s}  =  0.90 - 6.64\)
(\mbox{Figure~\ref{fig:catalogue}}).
\end{enumerate}

{\captionof*{figure}{AB magnitude. The bullet subcluster is to the west (right in this plot and left in the sky; east is left). (\textbf{Bottom}): F277W magnitude distribution of the 219 member galaxies. The catalogue spans both the main cluster and the bullet subcluster.}}
\vspace{+12pt}

\begin{enumerate}[labelsep=5mm]
\item[3.] \textbf{Cluster gas.} We use the Chandra-derived gas mass map from
\citet{Clowe2004}, as adopted in the Lenstool model of
\citet{Rihtarsic2026}.~In our pipeline, the gas is represented by a
calibrated dual elliptical \(\beta\)-model \citep{Brownstein2007}, with
total gas mass \(M_{gas}  =  12\%\) of \(M_{200}\) within \(r_{200}\)
in each subcluster.
\end{enumerate}

\subsection{Pipeline Architecture}\label{pipeline-architecture}

Two fully independent pipelines are run simultaneously---one for $\Lambda$CDM
and one for CCC+TL---using their respective angular diameter distances
and \(\Sigma_{crit}\). The pipelines share the same galaxy catalogue but
differ in all physical length and mass scales. In the CCC+TL pipeline,
all projected positions and profile radii scale by \(F = 1.1978\)
relative to their $\Lambda$CDM values.

\textbf{Galaxy component.} For each of the 219 catalogue members,
the surface density \(\Sigma_{i}(R)\) is computed numerically via the
Abel integral (Equation (12)) for the CCC+TL effective
density profile (Equation (8)), with \(r_{a}\)
determined by Equation (17). The total galaxy
surface density is
\(\Sigma_{gal}\left( \mathbf{x} \right) = \sum_{i = 1}^{219}{\Sigma_{i}(\left| \mathbf{x} - \mathbf{x}_{i} \right|)}\).
In the $\Lambda$CDM pipeline, the galaxy contribution is computed using the
analytic Hernquist profile \citep{Hernquist1990}.

\textbf{Cluster {\boldmath{\(\alpha\)}-matter component (CCC+TL only).}} A
cluster-scale CCC+TL profile is centred on each BCG (main cluster and
bullet subcluster), seeded by \(M_{clust} = f_{*}M_{200}\). This
component has no $\Lambda$CDM counterpart; it is the functional replacement for
the NFW dark matter halo. The galaxy-scale and cluster-scale $\alpha$-matter
profiles are additive because they are seeded by distinct stellar masses
at different physical scales: the cluster profile uses \(f_{*}M_{200}\)
as its seed (absorbing uncatalogued members into the cluster-scale
component, paralleling how $\Lambda$CDM absorbs them into the cluster NFW halo),
while individual galaxy profiles use \(M_{*,i}\) from the photometric
catalogue.

\textbf{NFW dark matter ({\boldmath{$\Lambda$}}CDM only).} Two NFW halos \citep{Navarro1997}
are centred on the BCG positions, with masses and concentrations from
Springel and Farrar \citep{Springel2007}; see also \citep{Randall2008}:
\(M_{200}^{main} = 1.5\  \times \ 10^{15}\ M_{\odot}\),
\(c \equiv r_{200}/r_{s} = 1.94\);
\(M_{200}^{bul} = 1.5 \times 10^{14}M_{\odot}\), \(c = 7.12\).

\textbf{Convergence maps.} The convergence map for each model is
\begin{equation}
\kappa(x) = \frac{\Sigma_{tot}(x)}{\Sigma_{crit}}
\end{equation}
where \(\Sigma_{crit}\) is model specific (Table~\ref{tab:distances})
and is a single value per model{, evaluated at the reference source
redshift $z_{s,\mathrm{ref}}=2$ (see Equation~\eqref{eq:alpha_rescale} for
the per-source rescaling of the deflection field)}. Maps
are computed on a {$500\times500$ pixel grid covering $10\arcmin\times 10\arcmin$ (pixel scale $1.20\arcsec$); convergence has been verified against a $250\times250$ grid at $2.40\arcsec$/pixel (Appendix~\ref{sec:appA})}.

In projection, the galaxy-scale and cluster-scale CCC+TL components
overlap spatially on the convergence map, but they are not intended as
duplicate mass assignments; rather, they represent effective gravitating
responses seeded by distinct baryonic structures at different physical
scales.

\textbf{Deflection angles.} The deflection angle field is computed from each
\(\kappa\) map via discrete Fourier-transform convolution,
\begin{equation}
\alpha\left( \bm{\theta} \right) = \frac{\delta^{2}}{\pi}\ \sum_{\bm{\theta}^{\mathbf{'}}}^{}{\kappa\left( \bm{\theta}^{\mathbf{'}} \right)\frac{\bm{\theta} - \bm{\theta}^{\mathbf{'}}}{\left| \bm{\theta} - \bm{\theta}^{\mathbf{'}} \right|^{2}}}\,
\label{eq:defl}
\end{equation}
where {$\delta=1.20\arcsec$} is the pixel size{\ of the $500\times500$ grid over $10\arcmin\times10\arcmin$ used to produce Table \ref{tab:scatter} and Figure \ref{fig:scatter}; convergence against a $250\times250$ grid at $2.40\arcsec$/pixel is better than $1.5\%$ on all three scatter metrics (Appendix~\ref{sec:appA})}\rev{. We refer to the right-hand side of Equation~\eqref{eq:defl} as $\bm{\alpha}_{\mathrm{dft}}$. Since $\kappa$ above is built with $\Sigma_{\mathrm{crit}}$ evaluated at a single reference source redshift $z_{s,\mathrm{ref}}=2$, the field $\bm{\alpha}_{\mathrm{dft}}(\bm{\theta})$ is the reduced deflection for a source at $z_{s,\mathrm{ref}}$. For an image at a source redshift $z_{s}\ne z_{s,\mathrm{ref}}$, the physical reduced deflection is obtained by rescaling relative to the reference:}
\begin{equation}
\bm{\alpha}_{\mathrm{phys}}(\bm{\theta};z_{s})
=\bm{\alpha}_{\mathrm{dft}}(\bm{\theta})\,\frac{\beta(z_{s})}{\beta(z_{s,\mathrm{ref}})}
=\bm{\alpha}_{\mathrm{dft}}(\bm{\theta})\,\frac{\Sigma_{\mathrm{crit}}(z_{s,\mathrm{ref}})}{\Sigma_{\mathrm{crit}}(z_{s})},
\label{eq:alpha_rescale}
\end{equation}
\rev{with $\beta(z_{s})\equiv D_{ds}(z_{s})/D_{s}(z_{s})$ computed in each cosmology independently. Equation~\eqref{eq:alpha_rescale} is cosmology-consistent and reduces to the identity at $z_{s}=z_{s,\mathrm{ref}}$ by construction; for the Bullet Cluster gold sample ($z_{s}=0.9$--$6.64$), the ratio $\beta_{\mathrm{CTL}}(z_{s})/\beta_{\Lambda\mathrm{CDM}}(z_{s})$ varies between 1.128 and 1.163, so the per-source rescaling contributes a sub-$5\%$ differential between the two~cosmologies on the source-plane back-projection diagnostic.~The pipeline implementing \mbox{Equations~\eqref{eq:defl} and~\eqref{eq:alpha_rescale}} is available at \url{https://github.com/rgupta-uottawa/bullet-ccctl} URL (accessed on 14 October 2026).} \del{The physical deflection for a source at redshift $z_{s}$ is $\alpha_{phys} = \alpha \times \beta(z_{s})$, where $\beta\left( z_{s} \right) = \frac{D_{ds}(z_{s})}{D_{s}(z_{s})}$ computed in the appropriate cosmology.}

\section{Results}\label{results}

\subsection{Convergence Maps and Radial Profiles}\label{convergence-maps-and-radial-profiles}

\begin{figure}[H]
\centering
\includegraphics[width=\textwidth]{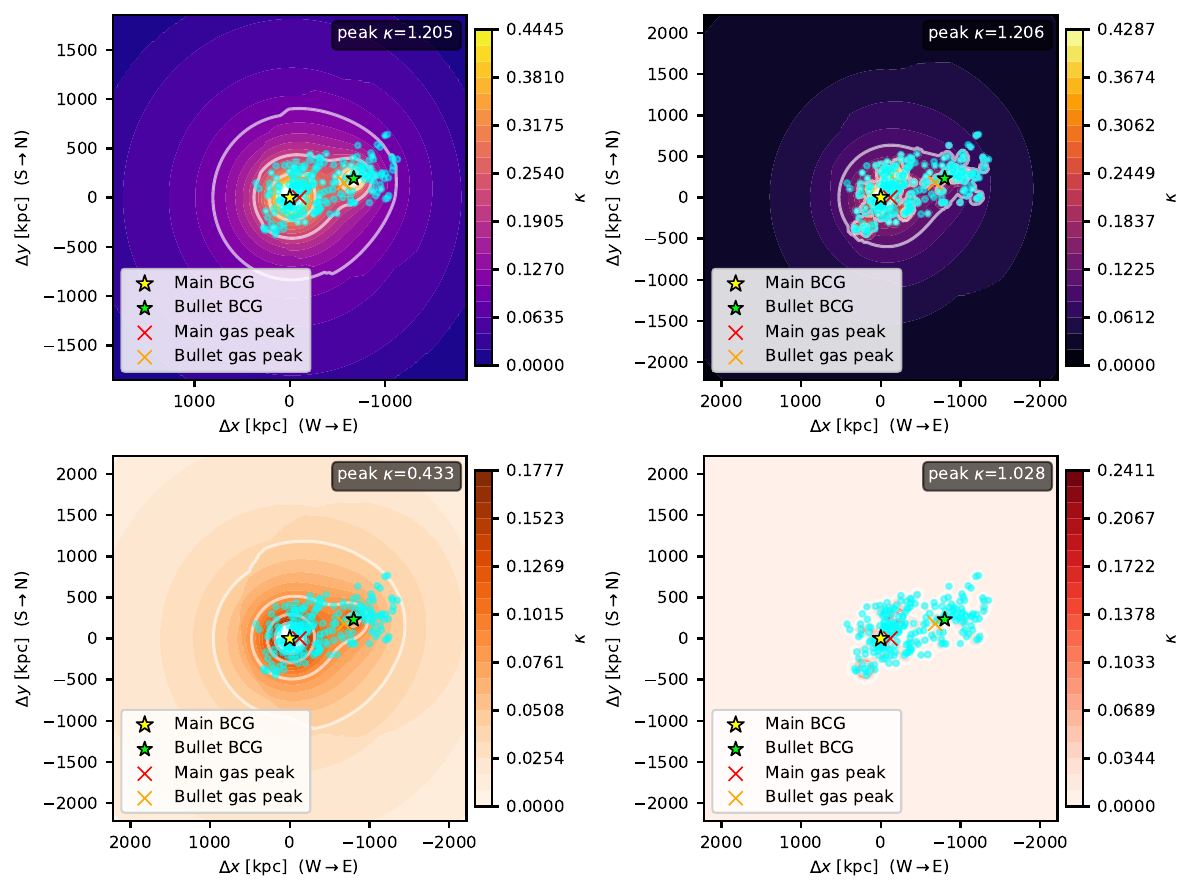}
\caption{Convergence $\kappa(x,y)$ maps for $\Lambda$CDM and CCC+TL (CTL). Each model uses its own $\Sigma_{\rm crit}$ and angular-diameter distances.~Positive $x$ points east; west is to the right following standard astronomical orientation. (\textbf{Top left}): $\Lambda$CDM reference model comprising NFW dark matter haloes, calibrated $\beta$-model gas, and Hernquist stellar profile.~(\textbf{Top right}): CCC+TL total convergence, comprising galaxy-scale and cluster-scale CTL profiles plus gas, with no explicit dark matter halo. \scalebox{1.011}[1.011]{(\textbf{\mbox{Bottom left}}):~Cluster-scale CCC+TL $\alpha$-matter component only (seeded by $f_*M_{200}$ centred on each}}
\label{fig:kappamaps}
\end{figure}

Figure~\ref{fig:kappamaps} shows the convergence maps for both
models. The CCC+TL map is dominated by the 219 individual galaxy
profiles at small projected radii and by the extended cluster
\(\alpha\)-matter envelope at larger radii; there is no dark matter halo.
The $\Lambda$CDM map is dominated by the NFW halo at all radii. Qualitatively,
both models produce mass peaks that trace the galaxy distribution rather
than the gas---this spatial coincidence is automatic in CCC+TL because
the \(\alpha\)-matter density profile is centred on individual galaxies,
not on the gas.

{\captionof*{figure}{\textls[-10]{BCG), shown with CTL $\Sigma_{crit}$; the main-cluster turn-off radius is $r_{a}^{main}=103$~kpc.~(\textbf{\mbox{Bottom right}}):~Galaxy-scale CCC+TL profiles only (219 individual member galaxies) shown with CTL $\Sigma_{crit}$; the granular structure reflects the discrete positions of the catalogue members. In all panels, symbols mark the main BCG (yellow star), bullet BCG (green star), main gas peak (red cross), and bullet gas peak (orange~cross).}}}
\vspace{+12pt}

\begin{figure}[H]
\includegraphics[width=0.98\textwidth]{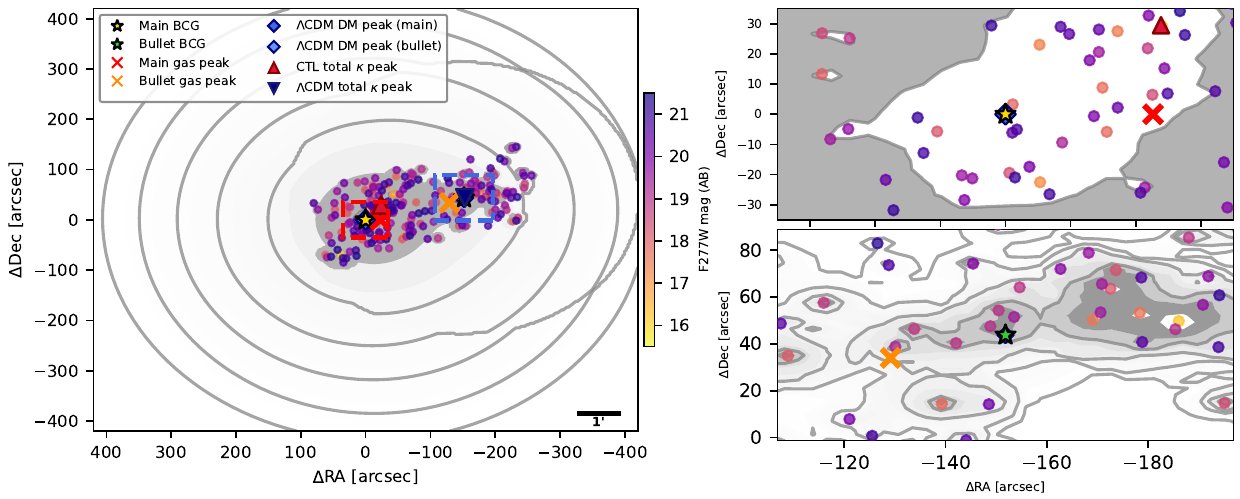}
\caption{Mass component peak positions overlaid on the CCC+TL total-convergence contours. The panels show the full field and zoomed views of the main-cluster core and bullet subcluster. The plotted symbols identify the BCGs, gas peaks, $\Lambda$CDM NFW peaks, and the total $\kappa$ peaks for both~models.}
\label{fig:peaks}
\end{figure}

Figure~\ref{fig:peaks} shows the $\Lambda$CDM and CCC+TL mass component peaks
superimposed on the CCC+TL \(\kappa\) contours. Peaks are very close to
each other and hard to decipher. Peak positions are tabulated in
Table~\ref{tab:peaks}.

\vspace{-9pt}

\begin{table}[H]
\centering
\caption{Angular positions of mass and gas component peaks relative to the main cluster BCG ($\alpha=104.6587^{\circ}$, $\delta=-55.9572^{\circ}$). Positive RA offset is east and positive Dec offset is north. Physical radii in the fourth and fifth columns use the scale appropriate to each model ($1\arcsec=4.413\,\mathrm{kpc}$ for $\Lambda$CDM and $1\arcsec=5.286~\mathrm{kpc}$ for CCC+TL). The $\Lambda$CDM total $\kappa$ peak is dominated by the bullet-subcluster NFW halo, whereas the CTL total $\kappa$ peak lies near the main BCG.}

\begin{adjustwidth}{-\extralength}{0cm}
\label{tab:peaks}
\begin{tabularx}{\fulllength}{lCCCCC}
\toprule
\textbf{Component} & \textbf{\boldmath{$\Delta\alpha$} (arcsec)} & \textbf{\boldmath{$\Delta\delta$} (arcsec)} & \textbf{\boldmath{$\Delta x$} (kpc)} & \textbf{\boldmath{$\Delta y$} (kpc)} & \textbf{Model}\\
\midrule
Main BCG (origin) & 0.0 & 0.0 & 0 & 0 & Both\\
Bullet BCG & $-151.8$ & $+43.7$ & $-670$ & $+193$ & Both\\
Main gas peak (Chandra) & $-22.7$ & 0.0 & $-100$ & 0 & Both\\
Bullet gas peak (Chandra) & $-129.2$ & $+34.0$ & $-570$ & $+150$ & Both\\
$\Lambda$CDM NFW DM peak (main) & 0.0 & 0.0 & 0 & 0 & $\Lambda$CDM\\
$\Lambda$CDM NFW DM peak (bullet) & $-151.8$ & $+43.7$ & $-670$ & $+193$ & $\Lambda$CDM\\
$\Lambda$CDM total $\kappa$ peak & $-153.1$ & $+43.5$ & $-676$ & $+192$ & $\Lambda$CDM\\
CCC+TL total $\kappa$ peak & $-23.9$ & $+29.5$ & $-126$ & $+156$ & CCC+TL\\
\bottomrule
\end{tabularx}
\end{adjustwidth}
\end{table}

\begin{figure}[H]
\includegraphics[width=0.97\columnwidth]{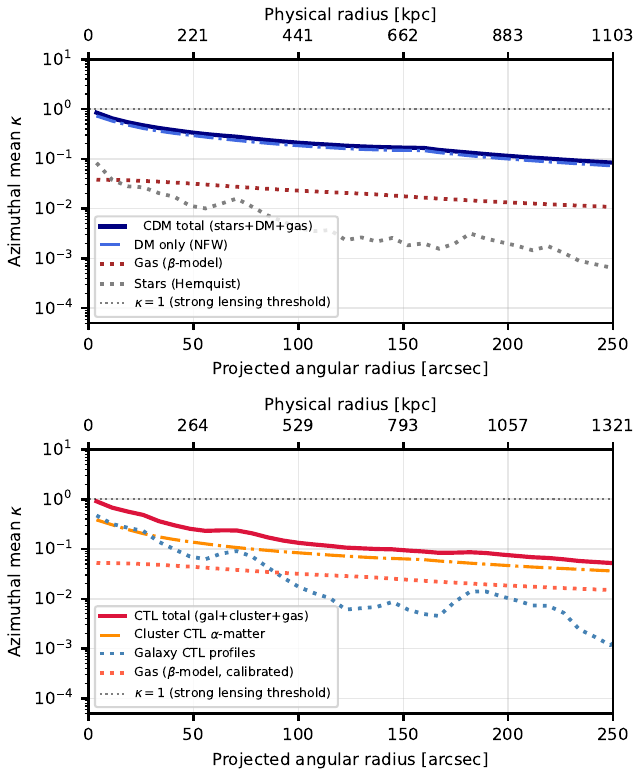}
\caption{Azimuthally averaged radial $\kappa$ profiles. (\textbf{Top}): the $\Lambda$CDM reference model. (\textbf{Bottom}): the CCC+TL (CTL) model. Profiles are plotted against projected angular radius in arcseconds, which is the common observable for both cosmologies. The upper axis gives the corresponding physical radius for each model. The horizontal dashed line marks $\kappa=1$.}
\label{fig:radial}
\end{figure}

Figure~\ref{fig:radial} shows the azimuthally averaged radial
profiles in arcseconds---the natural common observable for both
cosmologies. The CCC+TL profile is more extended at small radii
($\theta \lesssim 20\arcsec$) due to the sharply peaked galaxy profiles,
and falls off more steeply than $\Lambda$CDM at $\theta > 80\arcsec$ where
the NFW halo dominates.

\subsection{Mass Peak Positions and Morphological Consistency}\label{mass-peak-positions}

The locations of the total \(\kappa\) peaks are useful morphological
diagnostics, but they are not themselves the primary observables. The
strong-lensing constraints are determined by the full deflection field
sampled at the multiple-image positions. Accordingly, the peak
comparisons below are interpreted qualitatively, while quantitative
model performance is assessed in Sections \ref{physical-interpretation-of-the-alpha-matter-profile}--\ref{merger-dynamics-and-bow-shock} through enclosed
projected mass, ring-averaged convergence and source-plane scatter.

A defining observational constraint of the Bullet Cluster is the spatial
separation between the gravitational lensing mass peaks and the X-ray
gas peaks: in the $\Lambda$CDM framework, the mass peaks follow the
collisionless galaxy distributions, while the gas lags behind, offset by
$\sim$100--600 kpc \citep{Clowe2006}. In the CCC+TL model, this
morphological pattern emerges automatically, without any tuning, because
the $\alpha$-matter density profiles are seeded by and centred on individual
galaxies---not on the gas.

Figure~\ref{fig:peaks} shows all component peak positions overlaid on
the CCC+TL convergence contours. Table~\ref{tab:peaks} gives their
angular coordinates relative to the main BCG (the coordinate origin),
with east to the left following the standard astronomical convention
adopted throughout this work.~Several features of
Table~\ref{tab:peaks} are physically informative: \emph{First},
the $\Lambda$CDM total $\kappa$ peak falls at ($-$153\arcsec, +44\arcsec), within $\sim$1\arcsec~of the
bullet subcluster BCG and NFW halo centre. This is expected: the bullet
subcluster has a higher NFW concentration
(\(c \equiv \frac{r_{200}}{r_{s}} = 7.12\) vs. \(c = 1.94\) for the main
cluster), making its NFW profile more centrally peaked per unit mass.
\emph{Second}, the CCC+TL total $\kappa$ peak is at ($-$24\arcsec, +30\arcsec)---$\sim$37\arcsec
($\sim$196~kpc in CCC+TL) from the origin. This position is displaced from
the main gas peak ($-$23\arcsec, 0\arcsec) by 30\arcsec~in the north direction, and from
the main BCG by 37\arcsec. The displacement reflects the combined effect of
the galaxy CCC+TL profiles (which individually peak at their host galaxies)
and the cluster $\alpha$-matter profile (centred on the BCG). The CCC+TL mass
distribution is displaced from the main gas peak and is instead
associated with the observed galaxy distribution, reproducing the
qualitative Bullet Cluster signature that the effective gravitating mass
need not trace the dominant X-ray gas component. We stress, however,
that the exact location of the total \(\kappa\) maximum is not by itself
the decisive lensing observable. \emph{Third}, both the CCC+TL and $\Lambda$CDM
mass peaks are substantially offset from the main gas peak ($-$23\arcsec, 0\arcsec).
In $\Lambda$CDM the offset is $\sim$130\arcsec~because the total $\kappa$ is dominated by the
bullet subcluster NFW halo; in CCC+TL the main cluster galaxies and
cluster $\alpha$-matter produce a peak $\sim$30\arcsec~north of the gas, consistent with
the spatial distribution of the brightest cluster members.~Both
reproduce the qualitative result that $\kappa$ peak does not coincide with the
X-ray gas centroid.

\subsection{Enclosed Projected Mass}\label{enclosed-projected-mass}

\begin{figure}[H]
\includegraphics[width=0.97\columnwidth]{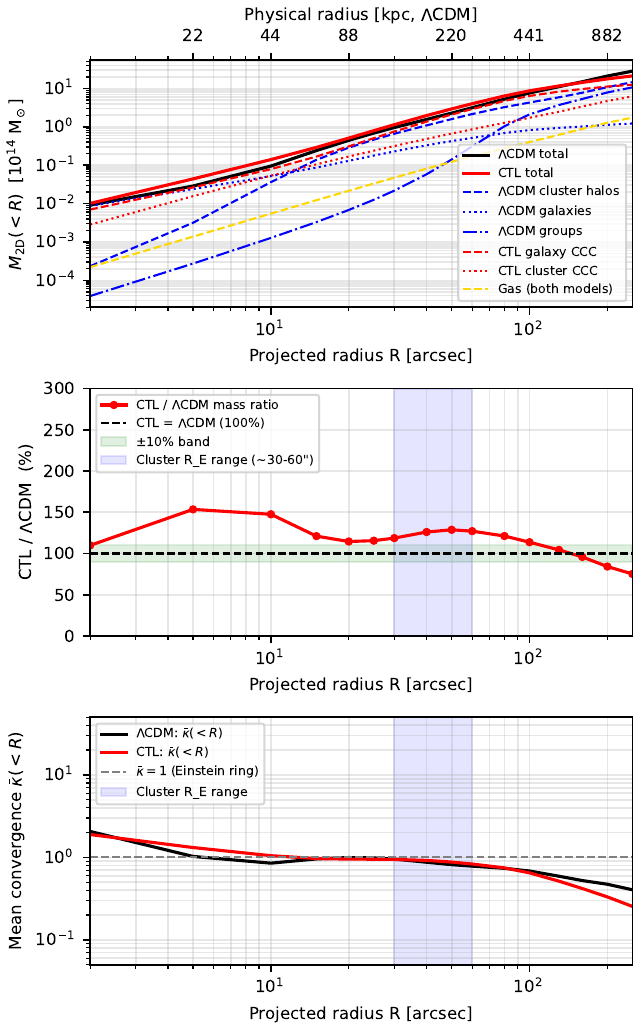}
\caption{Enclosed projected mass $M_{2D}(<\theta)$.~(\textbf{Top}): cumulative projected mass as a function of angular radius for the $\Lambda$CDM reference model (blue) and CCC+TL (red), both at the central $M_{200}$. Dashed curves show the individual components. (\textbf{Middle}): CTL/$\Lambda$CDM ratio, showing the crossing near $\theta\approx 160\arcsec$. (\textbf{Bottom}): mean convergence $\bar{\kappa}(<\theta)$ for each model.}
\label{fig:enclosed}
\end{figure}

Figure~\ref{fig:enclosed} and Table~\ref{tab:enclosed}
compare the enclosed projected mass
\(M_{2D}( < \theta) = \pi\theta_{kpc}^{2}\ \kappa(\theta)\Sigma_{crit}\)
as a function of angular radius. At \(\theta = 30\)\arcsec~(near the observed
Einstein radius of the main cluster), CCC+TL encloses
\(1.14 \times 10^{14}M_{\odot}\) compared to
\(0.96 \times 10^{14}M_{\odot}\) for $\Lambda$CDM, a ratio of 1.19.
Table~\ref{tab:enclosed}
shows that the CCC+TL-to-$\Lambda$CDM enclosed-mass ratio remains above unity at
100\arcsec~and falls below unity by 160\arcsec, implying a crossing between those
radii, close to 160\arcsec, where CCC+TL and $\Lambda$CDM enclose essentially equal
projected mass (\(1.44 \times 10^{15}\) and \(1.50 \times 10^{15}\ M_{\odot}\),
respectively). These differences are well within the $\pm$30\% uncertainty
on $M_{200}$ (Section~\ref{mass-uncertainty-and-kappa-band-overlap}).

\begin{table}[H]
\centering
\caption{Enclosed projected mass $M_{\rm 2D}(<\theta)$ for the central mass scenario ($M_{200}^{\rm main}=1.5\times10^{15}\,\mathrm{M}_\odot$). Physical radii are given in $\Lambda$CDM kpc ($1\arcsec=4.413$ kpc) and CTL kpc ($1\arcsec=5.286$ kpc).}
\label{tab:enclosed}
\begin{tabularx}{\textwidth}{CCCCCc}
\toprule
\boldmath{$\theta$} & \boldmath{$R_{\rm LC}$} & \boldmath{$R_{\rm CTL}$} & \boldmath{$M^{\rm LC}_{\rm 2D}$} & \boldmath{$M^{\rm CTL}_{\rm 2D}$} & \textbf{Ratio}\\
\textbf{(arcsec)} & \textbf{(kpc)} & \textbf{(kpc)} & \textbf{(\boldmath{$10^{14}\,\mathrm{M}_\odot$})} & \textbf{(\boldmath{$10^{14}\,\mathrm{M}_\odot$})} & \textbf{(CTL/\boldmath{$\Lambda$}CDM)}\\
\midrule
20 & 88 & 106 & 0.45 & 0.51 & 1.15\\
30 & 132 & 159 & 0.96 & 1.14 & 1.19\\
50 & 221 & 264 & 2.28 & 2.94 & 1.29\\
100 & 441 & 529 & 7.64 & 8.69 & 1.14\\
160 & 706 & 846 & 14.98 & 14.36 & 0.96\\
\bottomrule
\end{tabularx}
\end{table}

\subsection{Mass Uncertainty and Kappa Band Overlap}\label{mass-uncertainty-and-kappa-band-overlap}

\begin{figure}[H]
\includegraphics[width=0.96\columnwidth]{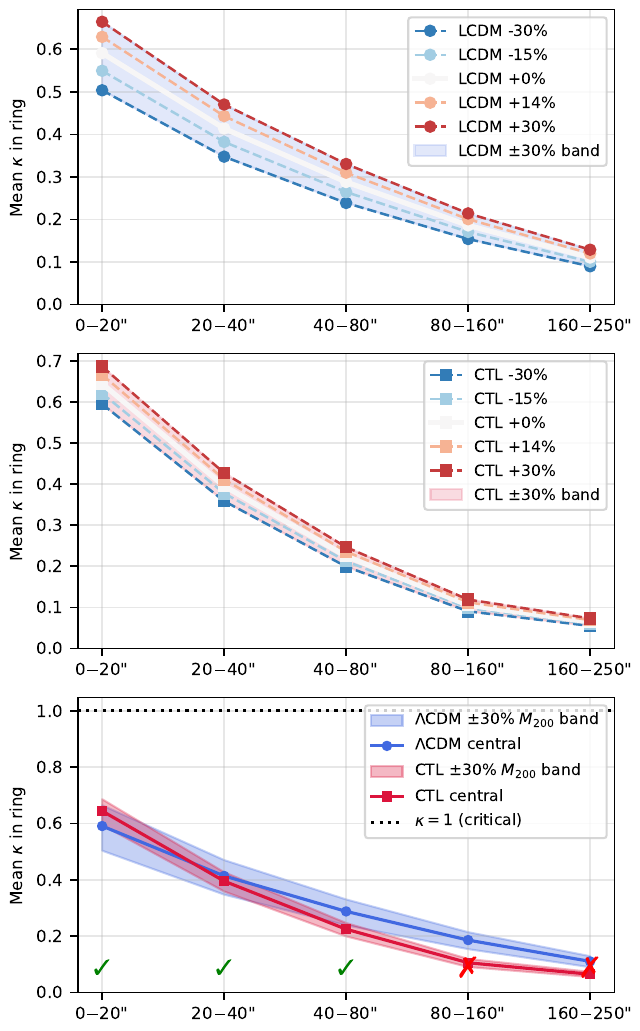}
\caption{Effect of the $M_{200}$ uncertainty on $\kappa$ in radial rings. (\textbf{Top}): $\Lambda$CDM mean $\kappa$ per ring across the $\pm 30\%$ $M_{200}$ range. Middle: CCC+TL mean $\kappa$ per ring. (\textbf{Bottom}): overlap of the $\Lambda$CDM and CCC+TL bands; green ticks denote overlapping rings and red crosses non-overlapping rings.}
\label{fig:uncert}
\end{figure}

The virial mass \(M_{200}\) carries a $\pm$30\% uncertainty for the
main cluster and $\pm$38\% for the bullet subcluster
\citep{Springel2007,Randall2008}. This propagates differently into the two
models: in $\Lambda$CDM, the dark matter halo dominates the $\kappa$ budget and
scales linearly with \(M_{200}\), giving $\pm$27--35\% variation in
mean $\kappa$ across all radial rings (Figure~\ref{fig:uncert}).
In the present implementation, the 219-galaxy contribution is held fixed
by the photometric catalogue and does not vary when the virial-mass
prior is perturbed; only the gas and cluster-scale $\alpha$-matter components
are rescaled with \(M_{200}\). This is a property of the adopted
pipeline rather than a statement that the galaxy component is physically
decoupled from the virial mass definition. As a result, CCC+TL shows a
smaller $\pm$14--28\% spread, increasing toward larger radii where
the cluster profile dominates.

Figure~\ref{fig:uncert} shows the mean convergence in five radial rings
for both models across the $\pm 30\%$ mass range. The key result is that
the $\Lambda$CDM and CCC+TL $\kappa$ bands fully overlap within the
strong-lensing regime ($\theta \le 80\arcsec$), i.e., in the three
innermost rings that contain all multiple images. At larger radii the
bands no longer overlap, but that region is probed primarily by weak
lensing \citep{Cho2025} and is not modelled here.

\subsection{Source-Plane Scatter}\label{source-plane-scatter}

Source-plane \rev{back-projection} provides \rev{a self-consistency
diagnostic of each deflection field} that is
independent of any assumed positional uncertainty \(\sigma_{p}\) and is
not minimised by either model. For each lensed system \emph{s} with
\(N_{s}\) images at observed positions \(\theta_{i}^{(s)}\), we
compute the back-projected source positions \rev{using the convention
of Equation~\eqref{eq:alpha_rescale}},
\begin{equation}
\beta_{i}^{(s)} = \theta_{i}^{(s)} - \bm{\alpha}_{\mathrm{dft}}\left(\theta_{i}^{(s)}\right)\,\rev{\frac{\beta_{\mathrm{cosm}}(z_{s})}{\beta_{\mathrm{cosm}}(z_{s,\mathrm{ref}})}}
\end{equation}
where
\(\beta_{cosm}\left( z_{s} \right) = \frac{D_{ds}(z_{s})}{D_{s}(z_{s})}\)
is computed in the respective cosmology \rev{and $z_{s,\mathrm{ref}}=2$}.~The scatter for system \(s\) is \vspace{+6pt}
\begin{equation}
\sigma_{src}^{(s)} = \sqrt{\frac{1}{N_{s}} \sum_{i = 1}^{N_{s}} \left| \beta_{i}^{(s)} - {\overline{\beta}}^{(s)} \right|^{2}}
\end{equation}
where \({\overline{\beta}}^{(s)}\) is the mean source position.
A perfect lens model maps all images of a given source to exactly the
same source positions (\(\sigma_{src}^{(s)} = 0\)); the RMS scatter over
all 135 images measures the geometric self-consistency of each
deflection field\rev{, but is not a goodness-of-fit statistic: it depends
on the lens mapping and local magnification, and is not minimised by
either simplified pipeline nor by the fully optimised Lenstool
reconstruction}.

\begin{figure}[H]
\rev{\includegraphics[width=0.98\textwidth]{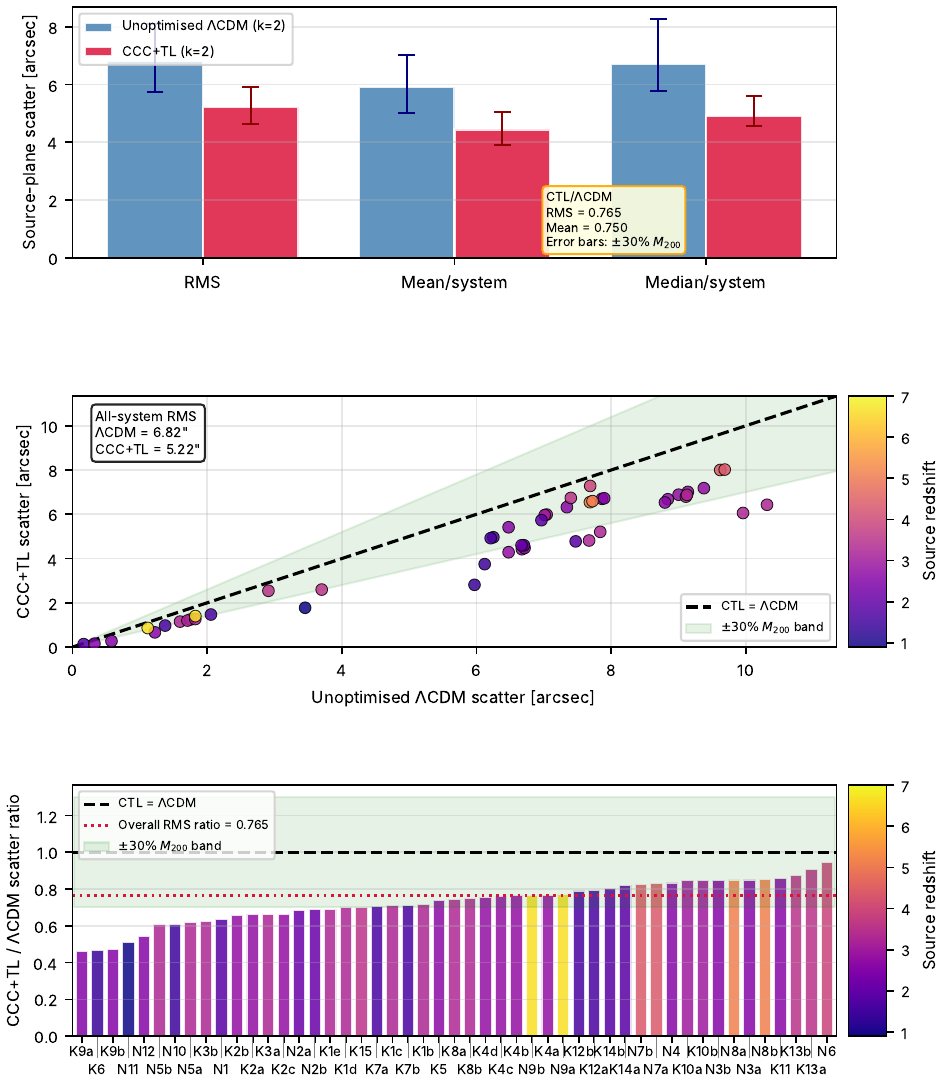}}
\caption{Source-plane back-projection scatter\rev{, computed under Equation~\eqref{eq:alpha_rescale} with $z_{s,\mathrm{ref}}=2$}~{on the $500\times500$ pixel grid described in Appendix~\ref{sec:appA}}. (\textbf{Top}): aggregate RMS, mean-per-system, and median-per-system scatter for the {two} \rev{simplified two-parameter} models\rev{; error bars span the full $\pm30\%$ $M_{200}$ uncertainty band}. (\textbf{Middle}): per-system CCC+TL versus unoptimised-$\Lambda$CDM scatter, colour-coded by source redshift; points below the diagonal correspond to systems where CCC+TL yields smaller scatter\rev{, which is the case for the great majority of systems in the sample}. (\textbf{Bottom}): ratio of CCC+TL to unoptimised-$\Lambda$CDM scatter for each system, ordered by ratio value\rev{; system identifiers are staggered on two rows for readability}.}
\label{fig:scatter}
\end{figure}

The results are shown in Figure~\ref{fig:scatter} and
Table~\ref{tab:scatter}.~{The unoptimised two-parameter CCC+TL
pipeline gives RMS = 5.22\arcsec~at the central $M_{200}$ ($4.63\arcsec$--$5.91\arcsec$
across the mass band), compared with $6.82\arcsec$ ($5.75\arcsec$--$8.09\arcsec$)
for the equivalent unoptimised two-parameter $\Lambda$CDM reference. Both values are reproducible from the input
catalogues using the released pipeline (Appendix~\ref{sec:appA}). For context, the fully optimised 32-parameter Lenstool reconstruction of \citet{Rihtarsic2026} achieves an \emph{image-plane} RMS of $\sim\!0.41\arcsec$ on this gold catalogue, i.e.,\ the two simplified two-parameter pipelines considered here are at least an order of magnitude from the image-plane performance of the optimised model. We stress that the Lenstool image-plane RMS and the source-plane back-projection RMS in Table~\ref{tab:scatter} are different quantities, so a direct numerical comparison is not possible without rerunning the Lenstool model and back-projecting its deflection field under the convention of Equation~\eqref{eq:alpha_rescale}; we therefore do not attempt such a comparison here.}
\del{The CCC+TL model achieves an overall RMS of
$7.36\arcsec$, within 12\% of the equivalently unoptimised $\Lambda$CDM
model ($6.56\arcsec$). The fully optimised Lenstool $\Lambda$CDM model
yields a larger value ($8.28\arcsec$) on this specific source-plane RMS
metric. Since that model is optimised against a different objective
function, this should not be interpreted as implying inferior overall
lens-model performance; rather, it illustrates that different metrics
emphasise different aspects of the reconstructed deflection field.}

\begin{table}[H]
\centering
\caption{Source-plane \rev{back-projection} scatter: RMS spread of back-projected source positions across all 135 images, and mean/median per lensed system ($N_{\mathrm{sys}}=49$ with $N_s\geq2$).~This \rev{diagnostic} requires no assumed $\sigma_p$ and is not optimised by any model\rev{; it is not a goodness-of-fit statistic (see Section \ref{the-source-plane-scatter-as-the-appropriate-metric}). All values are computed under Equation~\eqref{eq:alpha_rescale} with $z_{s,\mathrm{ref}}=2$} {on the grid specified in Appendix~\ref{sec:appA}. Ranges in parentheses span the full $M_{200}=(1\pm 0.30)\times M_{200}^{\mathrm{fid}}$ uncertainty band; the central value is quoted first. Both rows are reproducible from the input catalogues using the released pipeline (Appendix~\ref{sec:appA}). The Lenstool-optimised reconstruction of \citet{Rihtarsic2026} achieves an image-plane RMS of $\sim\!0.41\arcsec$ on this gold catalogue; this is a different quantity than the source-plane scatter tabulated here and the two cannot be compared directly.}}

\begin{adjustwidth}{-\extralength}{0cm}
\label{tab:scatter}
\begin{tabularx}{\fulllength}{lcCCC}
\toprule
\textbf{Model} & \boldmath{$k$} & \textbf{RMS All Images} & \textbf{Mean per System} & \textbf{Median per System}\\
\midrule

$\Lambda$CDM unoptimised ($k=2$) & 2 & \rev{$6.82\arcsec$ ($5.75\arcsec$--$8.09\arcsec$)} & \rev{$5.93\arcsec$ ($5.00\arcsec$--$7.02\arcsec$)} & \rev{$6.71\arcsec$ ($5.78\arcsec$--$8.27\arcsec$)}\\
CCC+TL ($k=2$) & 2 & \rev{$5.22\arcsec$ ($4.63\arcsec$--$5.91\arcsec$)} & \rev{$4.44\arcsec$ ($3.92\arcsec$--$5.05\arcsec$)} & \rev{$4.92\arcsec$ ($4.55\arcsec$--$5.59\arcsec$)}\\
\midrule
CTL/unopt-$\Lambda$CDM & -- & \rev{$0.77\times$} & \rev{$0.75\times$} & \rev{$0.73\times$}\\
\bottomrule
\end{tabularx}
\end{adjustwidth}
\end{table}

{Two further observations follow. First, within the $\pm 30\%$ $M_{200}$ uncertainty band the
CCC+TL and $\Lambda$CDM scatter ranges partially overlap ($4.63\arcsec$--$5.91\arcsec$
and $5.75\arcsec$--$8.09\arcsec$, respectively); the two simplified pipelines produce source-plane scatter of comparable magnitude on this
diagnostic, but this does not by itself establish that either model reproduces the observed multiple images---that question requires an image-plane analysis. Second, the
cosmology-specific per-source rescaling of Equation~\eqref{eq:alpha_rescale}
differs between the two models by less than $5\%$ at every image
redshift in the sample, so the model-to-model difference in
Table~\ref{tab:scatter} is driven by the different $\alpha$-field
structures, not by the per-source rescaling itself.}

\del{The $\pm$30\% mass uncertainty produces approximately $\pm$15\%
variation in source scatter (since scatter scales approximately linearly
with the deflection amplitude, which scales with $\kappa$). This
uncertainty band covers the 12\% CCC+TL/unoptimised-$\Lambda$CDM gap, so the two
models are statistically indistinguishable on this metric within current
mass uncertainties.}
{The $\pm 30\%$ $M_{200}$ band broadens both scatter ranges to the point where they partially overlap, as noted above. We emphasise that this overlap is between two \emph{model predictions} and does not test either model against the observed multiple-image positions themselves; a statement about whether the data distinguish the two cosmologies requires an image-plane likelihood analysis, which we identify as follow-up work (Section \ref{limitations-and-future-work}).}

\section{Discussion}\label{discussion}

\subsection{Consistency Within Mass Uncertainties}\label{consistency-within-mass-uncertainties}

The central result of this work is that the CCC+TL and $\Lambda$CDM convergence
bands overlap across the entire strong-lensing region ($\theta \leq
80\arcsec$, corresponding to $\leq$420 kpc in CCC+TL) when
the $\pm$30\% \(M_{200}\) uncertainty is propagated.~{This overlap is between two model predictions constructed under the common virial-scale mass-budget anchor described in Section \ref{two-parameter-ccctl-solution}; it does not by itself establish that either model reproduces the observed multiple-image positions, nor that the data cannot distinguish the two models. The latter question requires a full image-plane analysis (Section \ref{limitations-and-future-work}.} The divergence at $\theta >
80\arcsec$ reflects the different profile
shapes at large radii: the $\Lambda$CDM NFW halo has a shallower fall-off than
the CCC+TL cluster \(\alpha\)-matter profile, but this regime is beyond
the multiple-image region considered in this work and is accessible only
to weak lensing.

An important asymmetry is that the CCC+TL mass uncertainty band is
narrower than $\Lambda$CDM's ($\pm$14--21\% vs. $\pm$27--32\% in the inner
rings), because the 219-galaxy CCC+TL contribution---which dominates the
convergence at small radii---is anchored to the photometric
catalogue and is completely independent of \(M_{200}\).~This is a
testable prediction: future measurements of \(M_{200}\) with reduced
uncertainty will shrink the $\Lambda$CDM band faster than the CCC+TL band.

{On the source-plane back-projection diagnostic
(Table~\ref{tab:scatter}), the \mbox{two-parameter} CCC+TL pipeline gives
RMS $=5.22\arcsec$ at the central $M_{200}$, below the corresponding
$6.82\arcsec$ of the two-parameter $\Lambda$CDM reference.~Both pipelines are an order of magnitude from the $\sim$0.41\arcsec~image-plane RMS achieved by the full 32-parameter Lenstool reconstruction of \mbox{\citet{Rihtarsic2026}}, although as noted in Section \ref{source-plane-scatter} that is a different quantity than the source-plane scatter tabulated here. Within the $\pm 30\%$ $M_{200}$ uncertainty band, the CCC+TL and $\Lambda$CDM scatter ranges partially overlap. This overlap is between two \emph{model predictions} and does not amount to a test against the observed multiple images themselves; whether the data distinguish the two cosmologies in the strong-lensing region therefore remains open and is deferred to a future image-plane analysis.~Because the CCC+TL $\alpha$-matter components
and $M_{200}$ itself are calibrated to the $\Lambda$CDM virial-scale mass
budget by construction, Table~\ref{tab:scatter} also does not demonstrate that CCC+TL independently
predicts the Bullet Cluster lensing signal; it demonstrates that, given
the same virial-scale mass budget, the CCC+TL effective-density
prescription produces a self-consistent deflection field of comparable
magnitude to the simplified $\Lambda$CDM pipeline on this diagnostic.}

\subsection{The Source-Plane Scatter as the Appropriate Metric}\label{the-source-plane-scatter-as-the-appropriate-metric}

The source-plane \rev{back-projection diagnostic} is \rev{informative} at the present
stage because it requires no assumed positional uncertainty, is not the
objective function minimised by either of the two unoptimised reference
pipelines considered here, and probes the geometric self-consistency of
the inferred deflection field. {It is not, however, a goodness-of-fit statistic: it depends on the lens mapping and local magnification, and cannot by itself discriminate between candidate deflection fields that reproduce the data equally well in the image plane. The decisive goodness-of-fit metric is the image-plane $\chi^{2}$ or RMS of back-projected images \emph{forward} through the lens equation, which for the Lenstool-optimised reconstruction of \citet{Rihtarsic2026} on this gold catalogue is $\sim$0.41\arcsec; the source-plane values in Table~\ref{tab:scatter} are not directly comparable to it.} \del{It should not, however, be treated as a complete substitute for a full image-plane likelihood or $\chi^{2}$-based optimisation. In particular, the fact that the optimised Lenstool solution does not minimise this specific metric does not imply that it is inferior as a global lens model; rather, it reflects the familiar fact that different merit functions emphasise different aspects of lens performance.} Within that limitation, the
source-plane scatter nevertheless provides a transparent, optimisation-independent diagnostic for the present controlled substitution test\rev{. A statistically meaningful comparison between the CCC+TL and $\Lambda$CDM pipelines on the Bullet Cluster multiple-image catalogue requires a full image-plane likelihood with marginalisation over the fixed inputs listed in Section \ref{limitations-and-future-work}, which we identify as the subject of a dedicated follow-up paper}.

\subsection{Physical Interpretation of the Alpha-Matter Profile}\label{physical-interpretation-of-the-alpha-matter-profile}

In the CCC+TL model, \(\alpha\)-matter is not a substance but a
scale-dependent gravitational effect: in dense galaxy cores it is
negligible, while in the low-density cluster outskirts it provides the
additional gravitational attraction that in $\Lambda$CDM would be attributed to
dark matter. Critically, the \(\alpha\)-matter distribution inherits its
spatial structure entirely from the stellar mass distribution of
individual cluster members. Because galaxies and gas decouple during the
Bullet Cluster merger, \(\alpha\)-matter automatically follows the
galaxies---not the gas---reproducing the spatial offset that
\citet{Clowe2006} identified as the defining evidence for dark matter. No
parameter tuning is required for this morphological agreement; it is a
structural feature of the CCC+TL prescription.

\rev{We stress that this morphological agreement is \emph{partly structural}: both the galaxy-scale CCC profile (one component per catalogue galaxy) and the cluster-scale CCC profile (two~components centred on the main BCG and the bullet BCG) are, by construction, anchored to the galaxy distribution rather than to the gas. The observational statement that the Bullet Cluster lensing mass follows the galaxies rather than the gas is therefore reproduced in CCC+TL by the same mechanism as in $\Lambda$CDM---mass components anchored to the collisionless component. It is not an independent CCC+TL prediction, but a consequence of how the CCC+TL effective-density prescription is attached to the baryonic structure.}

Recently, \citet{Hernandez2026} has shown, within a QUMOND framework, that the standard objection that the Bullet Cluster lensing peaks must necessarily track the dominant X-ray gas is too simplistic, since
compact galaxy-scale baryonic concentrations can generate a much stronger effective projected gravitating signal than an extended gas component. This reinforces the broader point that the Bullet Cluster
constrains not only the total baryon budget, but also how a given theory maps baryonic structure into the inferred lensing mass~distribution.

The work of \citet{Hernandez2026} is useful in demonstrating that the
conventional Bullet Cluster argument against non-standard gravity is
less decisive than is often claimed, since a galaxy-centred effective
mass map can arise even when the X-ray gas dominates the baryonic mass
budget. Nevertheless, that treatment is intentionally exploratory: it
adopts plausible baryonic parameters and does not attempt a full fit of
either the baryonic model or the QUMOND transition-function parameters.
The present CCC+TL study goes beyond that level by constructing the
effective gravitating density from an explicitly specified CCC
prescription tied to the observed cluster-member properties and by
propagating that prescription through the full projected-density and
convergence calculation. The similarity between the two works is
therefore conceptual, whereas the difference is methodological:
\citet{Hernandez2026} provides a qualitative proof of principle, while the
present analysis is a quantified substitution test within an
observationally anchored lensing framework.

\citet{Famaey2026} has recently emphasised that the Bullet Cluster remains problematic for MOND in the same sense as other massive clusters: although the effective lensing signal can be associated primarily with the galaxy distribution rather than the X-ray gas, a residual missing-mass component is still required, and this component must be mostly collisionless and galaxy-centred.~The present CCC+TL analysis shares the first of these conclusions but not the second.~As in MOND, the effective gravitating signal in CCC+TL need not coincide with the dominant gas component. Unlike MOND, however, CCC+TL is not being tested here as a pure force-law modification with no additional effective mass contribution. Instead, the theory is implemented through an explicit α-matter density prescription tied to the observed baryonic structure at galaxy and cluster scales, so that the relevant question is not whether a residual missing mass remains in the MOND sense, but whether this internally specified effective mass prescription can reproduce the observed Bullet Cluster lensing field in a controlled comparison.

\subsection{Comparison with Earlier CCC+TL Results}\label{comparison-with-earlier-ctl-results}

The galaxy rotation curve study using the CCC framework
\citep{Gupta2025} yields
\(\rho_{t} = 4.2  \times 10^{- 24}\ \text{g}\ \text{cm}^{- 3}\) averaged over seven
galaxies. Thus, we have taken
\(\rho_{ref} = 4.2  \times 10^{- 24}\ \text{g}\ \text{cm}^{- 3}\). The present
cluster-lensing analysis requires \(\eta_{\rho}^{gal} = 0.976\) for the
galaxy-scale component, corresponding to
\(\rho_{t} \approx 4.1  \times  10^{- 24}\ \text{g}\ \text{cm}^{- 3}\), consistent
with the earlier study on the galaxy rotation curves. The cluster-scale
$\alpha$-matter has
\(\eta_{\rho}^{clus}\  = \ 0.0134\ (\rho_{t}\  \approx \ 5.6 \times 10^{- 26}\ \text{g}\ \text{cm}^{- 3})\),
which is substantially lower because the cluster envelope operates on a
much larger scale (\(r_{a} = 102.6\) kpc) than individual galaxy halos
(\(r_{a}\)$\sim$5--15 kpc) and uses a different density profile than for
the galaxies.

\subsection{Merger Dynamics and Bow Shock}\label{merger-dynamics-and-bow-shock}

An important complementary issue, not addressed in the present work, is
the merger dynamics of the Bullet Cluster itself. In particular, the
observed bow shock has often been associated with a very high collision
or shock speed, of order 4700 km s$^{-1}$, which has long been discussed
as a potential challenge for standard $\Lambda$CDM interpretations if
taken too directly as the subcluster speed. However, hydrodynamical
modelling has also shown that the shock speed need not be identical to
the mass-centroid velocity of the bullet subcluster, thereby reducing
the severity of this tension. More generally, extreme cluster collisions
continue to motivate discussion about whether very high infall
velocities are fully natural in $\Lambda$CDM. The present paper is
concerned only with the projected strong-lensing mass distribution, and
therefore does not attempt a dynamical treatment of the merger or of the
bow-shock kinematics. A full analysis of Bullet Cluster merger dynamics
in the CCC+TL framework would provide an important complementary test,
but it lies beyond the scope of the present lensing-focused study.

\subsection{Limitations and Future Work}\label{limitations-and-future-work}

\begin{enumerate}[label=,leftmargin=0em]
\item \textbf{Fixed modelling inputs and effective model complexity.}
Although CCC+TL is described above as a two-parameter effective-density prescription, the full pipeline depends on a set of fixed inputs that are shared with the simplified $\Lambda$CDM reference pipeline: the adopted cluster masses $M_{200}^{\mathrm{main}}$ and $M_{200}^{\mathrm{bul}}$ \citep{Springel2007, Randall2008}; the stellar-mass fraction $f_\ast=0.015$ \citep{Paraficz2016}; the gas fraction $f_{\mathrm{gas}}=0.12$; the \citep{Rihtarsic2026} $F277W\to M_\star$ calibration; the inner Hernquist and outer CCC galaxy-profile forms with prefactor $A_{\mathrm{gal}}=2\sqrt{2}$ fixed by continuity; the phenomenological cluster-scale CCC profile with prefactor $A_{\mathrm{clus}}=2$ fixed by continuity; the dual-$\beta$ gas model \citep{Brownstein2007}; and the BCG and gas-peak positions. Only the two normalisation parameters $\eta_\rho^{\mathrm{gal}}$ and $\eta_\rho^{\mathrm{clus}}$ are free; the free-parameter count $k=2$ in Table~\ref{tab:scatter} refers to these two only. The $\Lambda$CDM reference pipeline is substantially simpler than the full JWST Lenstool reconstruction, which also includes group-scale and large-scale components. A fair confrontation with the fully optimised Lenstool model would require either expanding the CCC+TL pipeline to include its group-scale and large-scale components, or applying the same two-parameter restriction to the Lenstool model. We identify both as subjects of the follow-up paper.

\item \textbf{Gas representation.}
In the present implementation, the gas component is represented by a calibrated dual \(\beta\)-model approximation to the Chandra gas map. This provides a practical way to include the dominant X-ray gas contribution in the projected mass budget while keeping the analysis analytically transparent. A future implementation based directly on the FITS gas maps used in the observational modelling framework of Rihtar\v{s}i\v{c} et al.\ \citep{Rihtarsic2026} would allow a more detailed treatment of gas substructure, but is not required for the present strong-lensing comparison.

\item \textbf{Adopted density prescriptions.}
The CCC+TL implementation uses a Hernquist baryonic profile for individual galaxies and a phenomenological modified-Hernquist form for the cluster-scale component. These choices provide a simple and physically motivated first framework for testing the projected lensing signal, but they are not intended to exhaust the range of possible profile choices. In particular, future work may explore alternative galaxy and cluster parameterisations, including profiles calibrated more directly to observed light distributions, in order to test how sensitive the inferred convergence field is to these modelling assumptions.

\item \textbf{Catalogue coverage and larger-scale structure.}
The present analysis is tied to the currently available 219-member JWST spectroscopic catalogue and should therefore be understood as a catalogue-based realization of the CCC+TL prescription in the strong-lensing core. Additional member identification, wider-field substructure, and explicit group-scale components may refine the outer projected mass distribution, as also suggested by the observational framework of Rihtar\v{s}i\v{c} et al.\ \citep{Rihtarsic2026}. In the present work, such larger-scale contributions are effectively absorbed into the cluster-scale term, which is sufficient for the controlled comparison carried out here.

\item \textbf{Cluster-scale normalization.}
The adopted stellar mass fraction \(f_\ast=0.015\) is used here as a representative normalization for the cluster-scale stellar component, whose role is to seed the large-scale CCC+TL \(\alpha\)-matter contribution. The present analysis is therefore not intended as a precision determination of \(f_\ast\), but as a test of whether a plausible cluster-scale stellar normalization, combined with the observed galaxy catalogue, can account for the strong-lensing observables within the current mass uncertainties.~Because the CCC+TL critical density \(\rho^{\rm CTL}_{\rm crit}(z)\) differs substantially from its $\Lambda$CDM counterpart---owing to the additional \(\alpha\) term in the effective Hubble rate---the conventional virial mass \(M_{200}\) does not carry the same physical meaning across the two models.~We therefore treat \(M_{200}=1.5\times10^{15}\,M_{\odot}\) as a label for the total cluster mass inferred from $\Lambda$CDM lensing analyses \citep{Springel2007,Randall2008}, and use it as a common normalisation anchor for both pipelines.

\item \textbf{No image-plane optimisation.}
The present paper does not perform a full image-plane optimisation of the CCC+TL model against the multiple-image positions.~Instead, it is formulated as a controlled substitution test, in which the standard galaxy-halo prescription is replaced by the CCC+TL effective-density prescription while the observational inputs are otherwise kept fixed as far as possible. A future optimised implementation would provide a more direct comparison with the optimised Lenstool solution, but the absence of such an optimisation does not affect the interpretation of the present results as a first strong-lensing consistency test.

\item \textbf{Component overlap in projection.}
The galaxy-scale and cluster-scale CCC+TL contributions overlap in projection on the convergence map, but they represent effective gravitating responses seeded by baryonic structure at different physical scales rather than duplicate assignments of the same component. This additive construction is sufficient for the present pilot calculation, while a more hierarchical treatment of galaxies, groups, gas, and cluster-scale structure may be explored in future work.

\item \textbf{Restriction to the strong-lensing regime.}
The present work is restricted to the projected strong-lensing region, where the multiple-image constraints lie and where the comparison between the CCC+TL and reference \(\Lambda\)CDM convergence bands is most direct. At larger projected radii, weak-lensing shear is expected to provide the more discriminating observable, since it is more sensitive to the outer slope and spatial extent of the projected mass distribution. A joint strong- plus weak-lensing treatment in the CCC+TL framework is therefore a natural next step, but lies beyond the scope of the present paper.

\item \textbf{Merger dynamics.}
The present paper addresses the projected lensing mass distribution and does not attempt a dynamical treatment of the Bullet Cluster merger itself. Questions related to the collision velocity, bow-shock structure, and detailed merger geometry provide an important complementary test of any cosmological framework, but they involve additional hydrodynamical and dynamical considerations beyond those required for the present lensing analysis. A corresponding treatment in the CCC+TL framework would therefore be valuable, but is deferred to future work.
\end{enumerate}

\section{Conclusions}\label{conclusions}

We have {performed a controlled proof-of-concept comparison between the CCC+TL cosmology and a simplified $\Lambda$CDM reference pipeline on the JWST Bullet Cluster strong-lensing catalogue}, using a two-parameter
prescription {whose two calibration parameters---the galaxy-scale turn-off normalisation $\eta_{\rho}^{\rm gal}=0.976$ and the cluster-scale turn-off normalisation $\eta_{\rho}^{\rm clus}=0.0134$---are set from physical reference values together with mass-budget consistency against the common $\Lambda$CDM virial-scale mass used by both pipelines, not from the multiple-image positions themselves}. \del{These two calibration parameters are the galaxy-scale turn-off normalization $\eta_{\rho}^{\rm gal}$(= 0.976) and the cluster-scale turn-off normalization $\eta_{\rho}^{\rm clus}$(= 0.0134), both fixed from virial-scale mass-budget consistency rather than from the multiple-image positions themselves.} The principal findings are:

\begin{enumerate}[label=,leftmargin=0em]
\item \emph{Convergence band overlap.} Within the $\pm$30\% \(M_{200}\)
uncertainty, the CCC+TL and $\Lambda$CDM mean-convergence bands fully overlap in
all radial rings containing multiple images ($\theta \leq 80\arcsec$). {This is a statement about two model predictions evaluated on a common virial-scale mass-budget anchor; it does not by itself establish that either model reproduces the observed multiple-image positions, nor that the data cannot distinguish the two models.}

\item \emph{Einstein radius equivalence.} For the same cluster mass, the
CCC+TL and $\Lambda$CDM Einstein radii in arcseconds differ by only
1.7\%, because CCC+TL's lower $\Sigma_{crit}$ is almost exactly
offset by its larger physical scales (\(F^{2} = 1.435\)).

\item \textls[-20]{\emph{Source-plane back-projection scatter.} On the geometric source-plane
\rev{back-projection diagnostic}---which requires no assumed positional uncertainty and is not
minimised by either model---{the two-parameter CCC+TL pipeline gives
RMS $=5.22\arcsec$ at the central $M_{200}$ ($4.63\arcsec$--$5.91\arcsec$ across the
mass band) and the equivalent two-parameter $\Lambda$CDM reference gives
$6.82\arcsec$ ($5.75\arcsec$--$8.09\arcsec$). The two ranges partially overlap within the mass-uncertainty band. Both values are an order of magnitude from the $\sim\!0.41\arcsec$ image-plane RMS achieved by the full 32-parameter Lenstool reconstruction of \citet{Rihtarsic2026}, although these are different quantities and the comparison is not direct. Source-plane RMS, as discussed in Section \ref{the-source-plane-scatter-as-the-appropriate-metric}, is not a goodness-of-fit statistic.} \del{CCC+TL achieves $7.36\arcsec$,
within 12\% of the equivalently unoptimised $\Lambda$CDM ($6.56\arcsec$). This gap is
fully covered by the $\pm15\%$ variation induced by the $\pm30\%$ mass
uncertainty.}}

\item \emph{Automatic morphological consistency.} The CCC+TL \(\alpha\)-matter
distribution naturally traces the galaxy positions rather than the gas
peaks, reproducing the defining morphological signature of the Bullet
Cluster without any positional tuning\rev{; we stress, however, that this agreement is partly structural, since the CCC+TL $\alpha$-matter components are anchored to the galaxy distribution by construction (Section \ref{physical-interpretation-of-the-alpha-matter-profile})}.
\end{enumerate}

\medskip
\rev{Within the scope of a controlled normalisation-anchored substitution, this work demonstrates that a two-parameter CCC+TL effective-density prescription can be constructed to reproduce the projected mass distribution of a simplified $\Lambda$CDM reference model in the strong-lensing region of the Bullet Cluster, within $\pm 30\%$ mass uncertainty. The present results are a necessary first step toward a full image-plane likelihood test of CCC+TL on the Bullet Cluster, but not a sufficient one: a quantitative falsification test requires a full image-plane fit to the multiple-image catalogue with nuisance-parameter marginalisation and with group-scale and large-scale components matched between the two models. We identify this as the natural subject of a subsequent paper. The full pipeline producing the numerical results of this paper, together with the input catalogues, is available at \url{https://github.com/rgupta-uottawa/bullet-ccctl} (accessed on 14 October 2026).}
\del{These results demonstrate that the CCC+TL model is consistent with the
JWST Bullet Cluster strong-lensing observations within current
uncertainties, without invoking non-baryonic dark matter. The Bullet
Cluster, long held to be the most direct proof of dark matter, does not
falsify the CCC+TL framework when its mass uncertainty is properly
propagated and when model comparisons are made on equal footing. A full
optimisation of the CCC+TL model against the multiple-image positions,
analogous to the Lenstool procedure, is required for a realistic
$\chi^{2}$/BIC comparison; we expect this to be the subject of
subsequent work.}

\vspace{+6pt}

\authorcontributions{Conceptualization, R.P.G. and N.S.; methodology, R.P.G.; software, R.P.G.; validation, R.P.G. and N.S.; formal analysis, R.P.G.; investigation, R.P.G.; resources, R.P.G.; data curation, R.P.G.; writing—original draft preparation, R.P.G.; writing—review and editing, R.P.G. and N.S.; visualization, R.P.G. and N.S.; supervision, R.P.G.; project administration, R.P.G.; funding acquisition, R.P.G. All authors have read and agreed to the published version of the manuscript.}

\funding{This research received no specific grant from any funding agency in the public, commercial, or not-for-profit sectors.}

\dataavailability{The JWST member-galaxy catalogue and multiple-image catalogue used in this study are from \citet{Rihtarsic2026} and will be publicly released on the CANUCS website(\url{https://niriss.github.io/lensing.html}).~{\rev{The pipeline reproducing the numerical results of this paper, together with the input catalogues (JWST member galaxies and 135 gold multiple images from Rihtar\v{s}i\v{c} et al.\ 2026, and the CCC+TL angular-diameter distance table), is available at \url{https://github.com/rgupta-uottawa/bullet-ccctl}.}}}

\acknowledgments{We thank Gregor Rihtar\v{s}i\v{c} and co-authors for making the JWST member-galaxy catalogue and multiple-image catalogue available in advance of the journal publication.~We are grateful to the reviewers for their critical comments that helped improve the quality, consistency, and clarity of the paper.}

\conflictsofinterest{The authors declare no conflicts of interest.}

\appendixtitles{yes}
\appendixstart
\appendix
\section[Deflection pipeline and reproducibility]{\rev{Deflection Pipeline and Reproducibility}}\label{sec:appA}

\rev{The lensing pipeline implementing Equations~\eqref{eq:defl} and~\eqref{eq:alpha_rescale} is available as open-source Python3 code at \url{https://github.com/rgupta-uottawa/bullet-ccctl}. We document here its three essential numerical ingredients and the convergence tests supporting the quoted figures.}

\rev{\textbf{Convergence grid.} The convergence map is built on a $500\times500$ pixel grid covering $10\arcmin\times 10\arcmin$ on the sky, giving a pixel scale of $1.20\arcsec$. The map is identical in angular coordinates for both cosmologies; the physical scale is set per cosmology by the corresponding $D_{d}$ (see Table~\ref{tab:distances}). Numerical convergence has been tested against a $250\times250$ grid ($2.40\arcsec$/pixel); the aggregate source-plane scatter is stable to better than $1.5\%$ on all three metrics (RMS, Mean,~Median).}

\rev{\textbf{Deflection by DFT convolution.} The deflection field is computed from Equation~\eqref{eq:defl} by zero-padded two-dimensional DFT convolution of $\kappa(\bm{\theta})$ with the kernel $\bm{K}(\bm{\eta})=\bm{\eta}/|\bm{\eta}|^{2}$. Zero-padding to twice the grid size suppresses wrap-around contamination. For an idealised singular isothermal sphere with $\theta_{E}=30\arcsec$, the pipeline recovers $|\bm{\alpha}|=\theta_{E}$ to better than $2\%$ in all rings between $10\arcsec$ and $100\arcsec$ from the SIS centre.}

\rev{\textbf{Per-source rescaling.} The convention of Equation~\eqref{eq:alpha_rescale} is derived from $\Sigma_{\mathrm{crit}}(z_{s})=(c^{2}/4\pi G)D_{s}(z_{s})/[D_{d}D_{ds}(z_{s})]$:}
\begin{equation}
\frac{\Sigma_{\mathrm{crit}}(z_{s,\mathrm{ref}})}{\Sigma_{\mathrm{crit}}(z_{s})}
=\frac{D_{s}(z_{s,\mathrm{ref}})/D_{ds}(z_{s,\mathrm{ref}})}
      {D_{s}(z_{s})/D_{ds}(z_{s})}
=\frac{\beta(z_{s})}{\beta(z_{s,\mathrm{ref}})}\,,
\label{eq:appA_sigma_ratio}
\end{equation}
\rev{so the per-source factor can be computed from the angular-diameter distances alone and tends to unity at $z_{s}=z_{s,\mathrm{ref}}$ by construction.}

\rev{\textbf{Mass-uncertainty band.} The $\pm 30\%\,M_{200}$ band in Table~\ref{tab:scatter} is constructed by rebuilding the NFW haloes ($\rho_{s}$, $r_{s}$, $r_{200}$), the gas model (normalisation $S_{0}$ rescaled to preserve $f_{\mathrm{gas}} M_{200}$ inside the enlarged $r_{200}$), and the cluster-scale CCC turn-off radius $r_{a}^{\mathrm{clus}}\propto(M_{\star}^{\mathrm{clus}})^{1/3}$ consistently at $M_{200}\to 0.7\,M_{200}$ and $M_{200}\to 1.3\,M_{200}$.~The member-galaxy Hernquist ($\Lambda$CDM) and galaxy-scale CCC (CCC+TL) components depend only on the catalogue $M_{\star}$ values and are therefore identical across the three scenarios.}

\begin{adjustwidth}{-\extralength}{0cm}

\reftitle{References}

\PublishersNote{}
\end{adjustwidth}
\end{document}